\documentclass[12pt,latin9,utf8]{article}
\usepackage[T1]{fontenc}
\usepackage[latin9]{inputenc}
\usepackage{graphicx}
\usepackage[unicode=true,pdfusetitle,
 bookmarks=true,bookmarksnumbered=false,bookmarksopen=false,
 breaklinks=false,pdfborder={0 0 1},backref=false,colorlinks=false]
 {hyperref}

\makeatletter
\newcommand{\pubnumber}{DPF2015-291}
\newcommand{\pubdate}{1 November 2015}

\usepackage{cite}

\def\Title#1{\begin{center} {\Large #1 } \end{center}}
\def\Author#1{\begin{center}{ \sc #1} \end{center}}
\def\Address#1{\begin{center}{ \it #1} \end{center}}

\newcommand{\pubblock}{\rightline{\begin{tabular}{l} \pubnumber\\
         \pubdate  \end{tabular}}}
\newenvironment{Abstract}{\begin{quotation}  }{\end{quotation}}
\newenvironment{Presented}{\begin{quotation} \begin{center} 
             PRESENTED AT\end{center}\bigskip 
      \begin{center}\begin{large}}{\end{large}\end{center} \end{quotation}}
\def\Acknowledgments{\bigskip  \bigskip \begin{center} \begin{large}
             \bf ACKNOWLEDGMENTS \end{large}\end{center}}

\makeatother

\begin{document}
\begin{titlepage} \pubblock

\vfill{}
\Title{Precision QCD for LHC Physics: The nCTEQ15 PDFs} \vfill{}
\Author{Fredrick I. Olness%
\footnote{Work supported in part by the U.S.~Department of Energy under grant
DE-FG02-04ER41299.%
} } \Address{ Southern Methodist University, Dallas, TX 75275, USA}

\vfill{}
\Address{ based on work in collaboration with } \Author{ D.~B.~Clark, 
E.~Godat,
T.~Ježo, 
C.~Keppel, 
K.~Kova\v{r}ík, 
A. Kusina, 
F.~Lyonnet, 
J.G.~Morf{\'{i}}n, 
P.~Nadolsky,
J.F.~Owens, 
I.~Schienbein,
J.Y.~Yu}

\vfill{}
\begin{Abstract} 

Searches for new physics at the LHC will increasingly depend on identifying
deviations from precision Standard Model (SM) predictions. At the
higher energy scales involved for the LHC Run 2, the heavy quarks
play a more prominent role than at the Tevatron. Recent theoretical
developments improve our ability to address multi-scale problems and
properly incorporate heavy quark masses across the full kinematic
range. These developments are incorporated into the new nCTEQ15 PDFs,
and we review these developments with respect to sample Run~2 measurements,
and identify areas where additional effort is required.

\end{Abstract} \vfill{}
\begin{Presented} DPF 2015\\
 The Meeting of the American Physical Society\\
 Division of Particles and Fields\\
 Ann Arbor, Michigan, August 4--8, 2015\\
 \end{Presented} \vfill{}
\end{titlepage}

\section{Introduction}

\nocite{Kovarik:2015cma, Kusina:2012vh, Kovarik:2010uv, Schienbein:2009kk}

\begin{figure}
\begin{centering}
\includegraphics[width=0.85\textwidth]{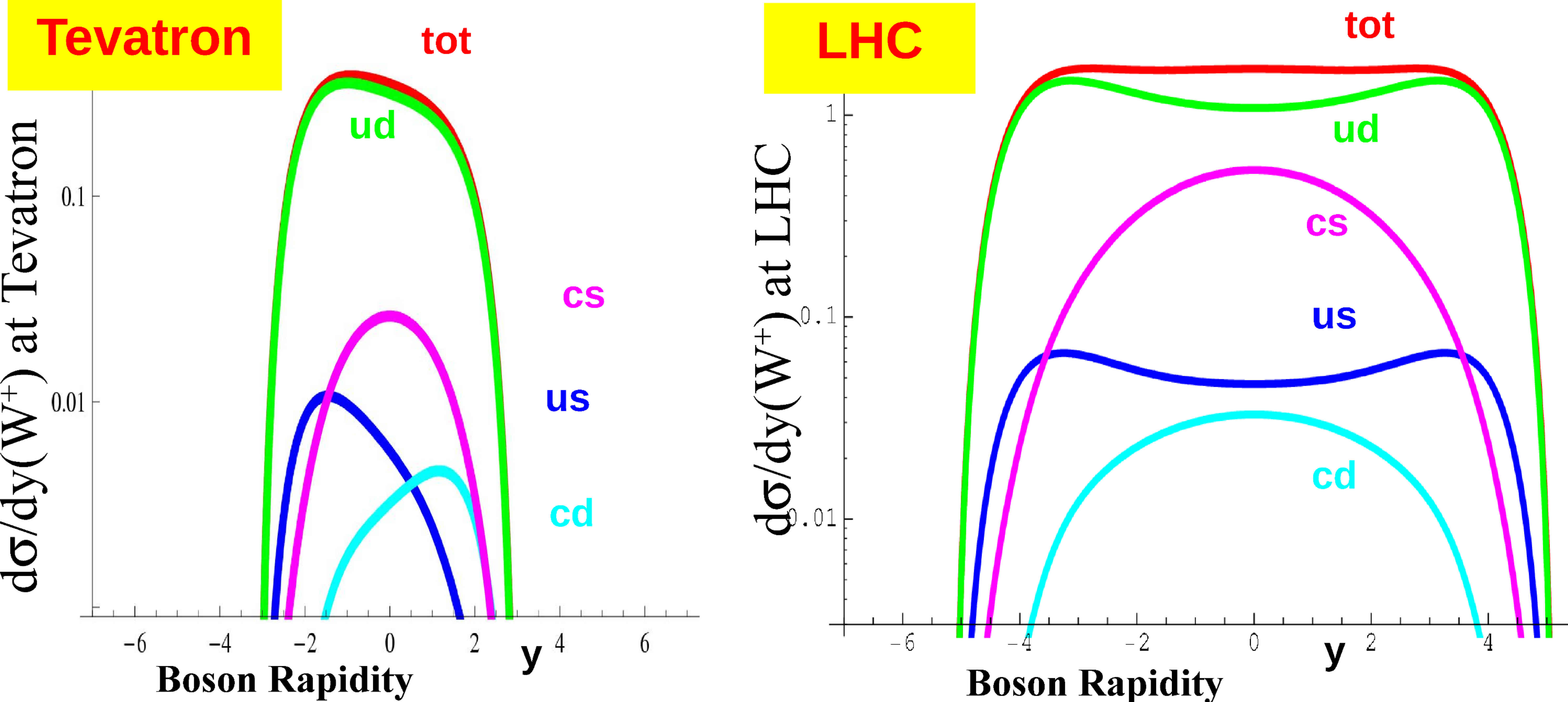}
\par\end{centering}

\caption{{\small{}The leading-order (LO) differential cross section ($d\sigma/dy$)
for $W^{+}$production at the Tevatron (2~TeV) and the LHC (14~TeV)
as a function of rapidity. The partonic contributions are also displayed
for $\{u\bar{d},c\bar{s},u\bar{s},c\bar{d}\}$. The vertical scales
are logarithmic. \label{fig:tevlhc}}}
\end{figure}

\begin{figure}
\centering{}\includegraphics[width=0.3\textwidth]{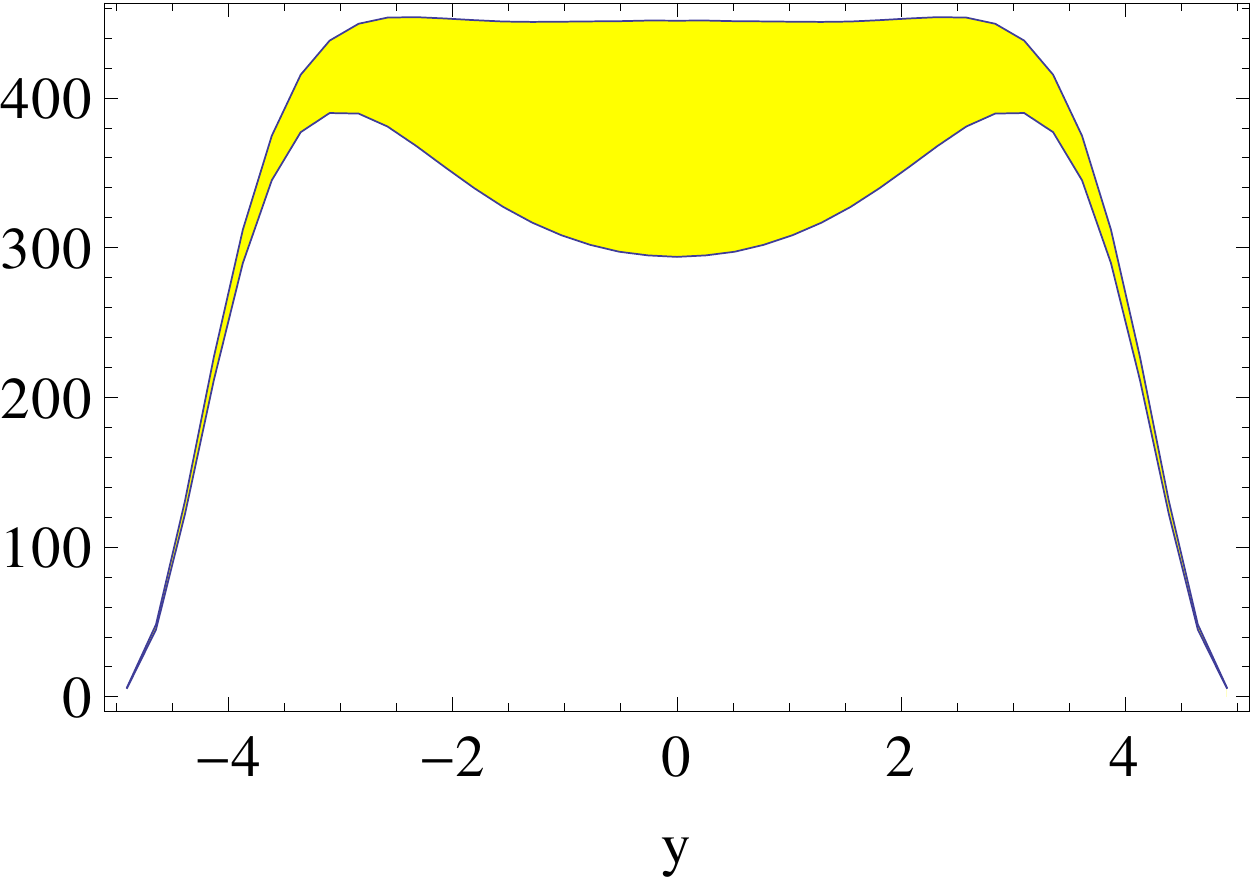}\hfil\includegraphics[width=0.3\textwidth]{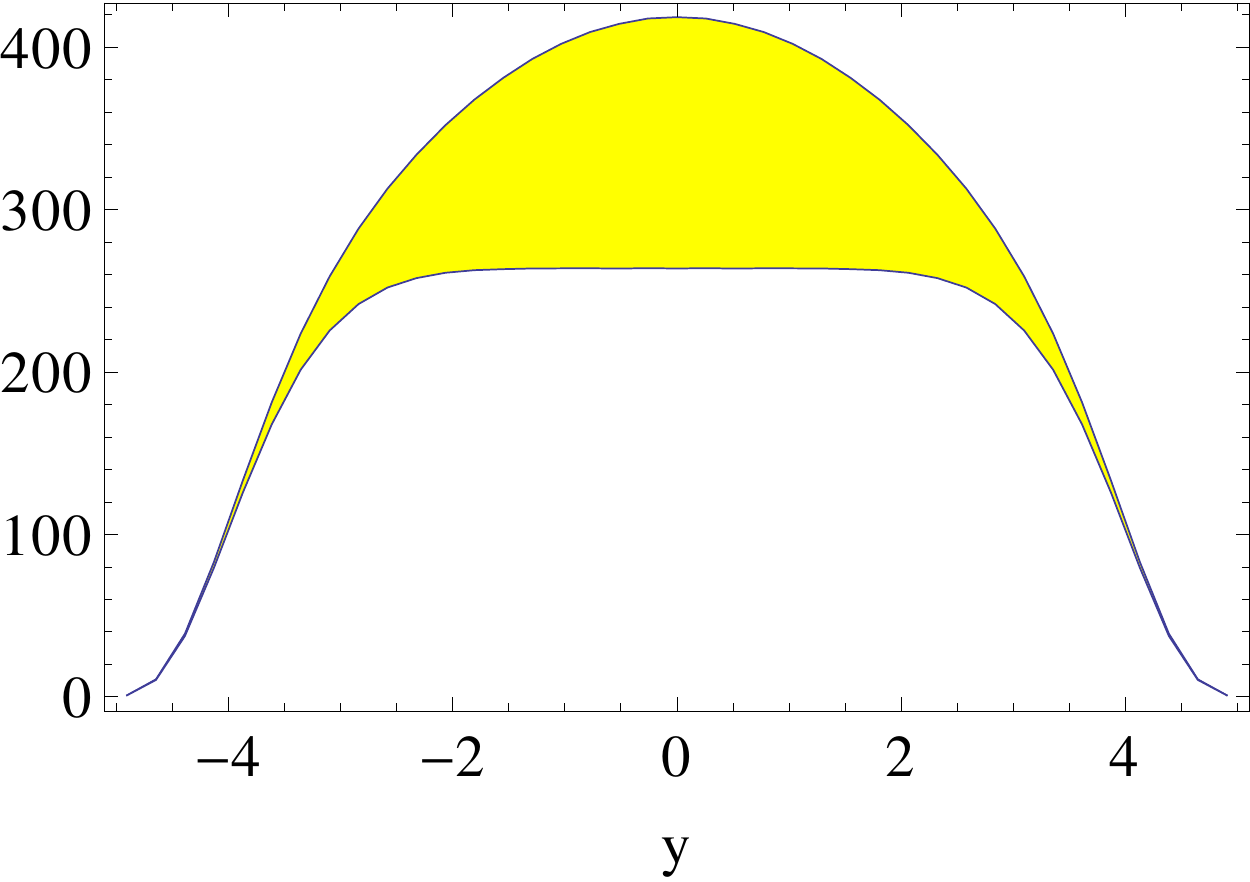}\hfil\includegraphics[width=0.3\textwidth]{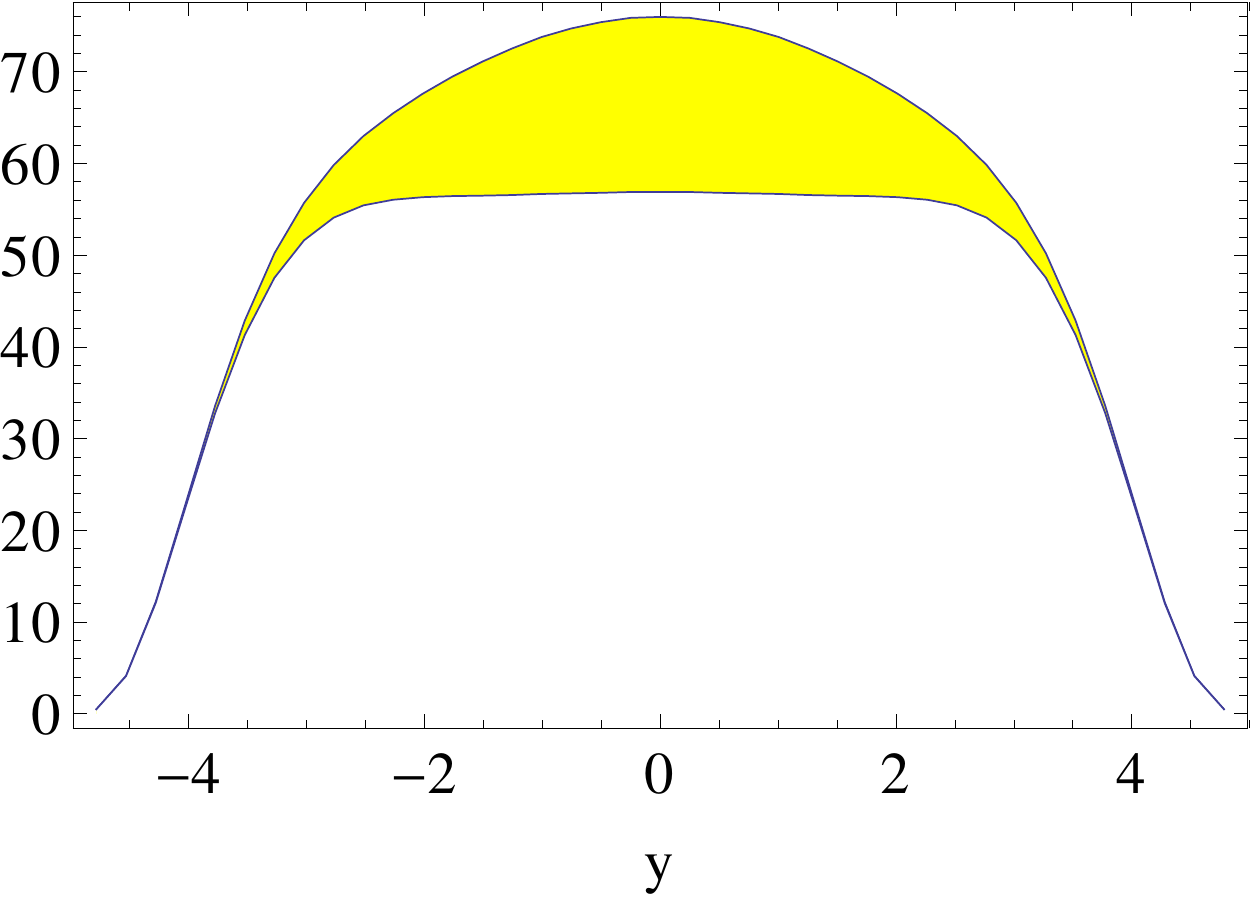}\caption{{\small{}The strange quark contribution (yellow) as a fraction of
the total $d^{2}\sigma/dM/dy$ in pb/GeV for $pp$ to $W^{+}$~(left),
$W^{-}$~(center), $Z$~(right) production at the LHC for 14~TeV
with CTEQ6.6 using the VRAP program at NNLO. C.f. Ref.~\cite{Kusina:2012vh}
for details. }\label{fig:strange}}
\end{figure}

Our field has seen major discoveries in recent years from a variety
of experiments, large and small, including a number recognized with
Nobel Prizes. The recent performance of the LHC has exceeded expectations
and produced an unprecedented number of events to be analyzed. On
the Intensity Frontier, Fermilab is advancing a number of high-precision
experiments (Muon $g-2$, Mu2e), as well as expanding its neutrino
program. Thus, there is a wealth of data to explore, and a comprehensive
analysis requires the most advanced and innovative tools.

As the accuracy of the experimental measurements increases, it is
essential to improve the theoretical calculations to match. If we
can make detailed predictions of W/Z/Higgs production (for example),
then we have the ability to distinguish a ``new physics'' signal
from an uncertain SM background process. To determine if the newly
discovered Higgs boson is that of the Standard Model (SM) or a more
exotic type, we must study both the production cross section and various
decay channels to make its proper characterization. In a complementary
manner, the Fermilab high-intensity high-statistics experiments force
us to reexamine previous assumptions (nuclear corrections, isospin
and lepton-flavor symmetries) and require us to extend our calculations
to increasingly high orders including subtle electroweak corrections.
The key step for all the above analyses is to make accurate predictions,
including realistic estimates of the underlying theoretical uncertainty.
The PDFs are at the heart of this program.

\section{PDF Flavor Determination \& Heavy Targets}

\begin{figure}
\begin{centering}
\includegraphics[bb=0bp 0bp 288bp 252bp,clip,width=0.35\textwidth]{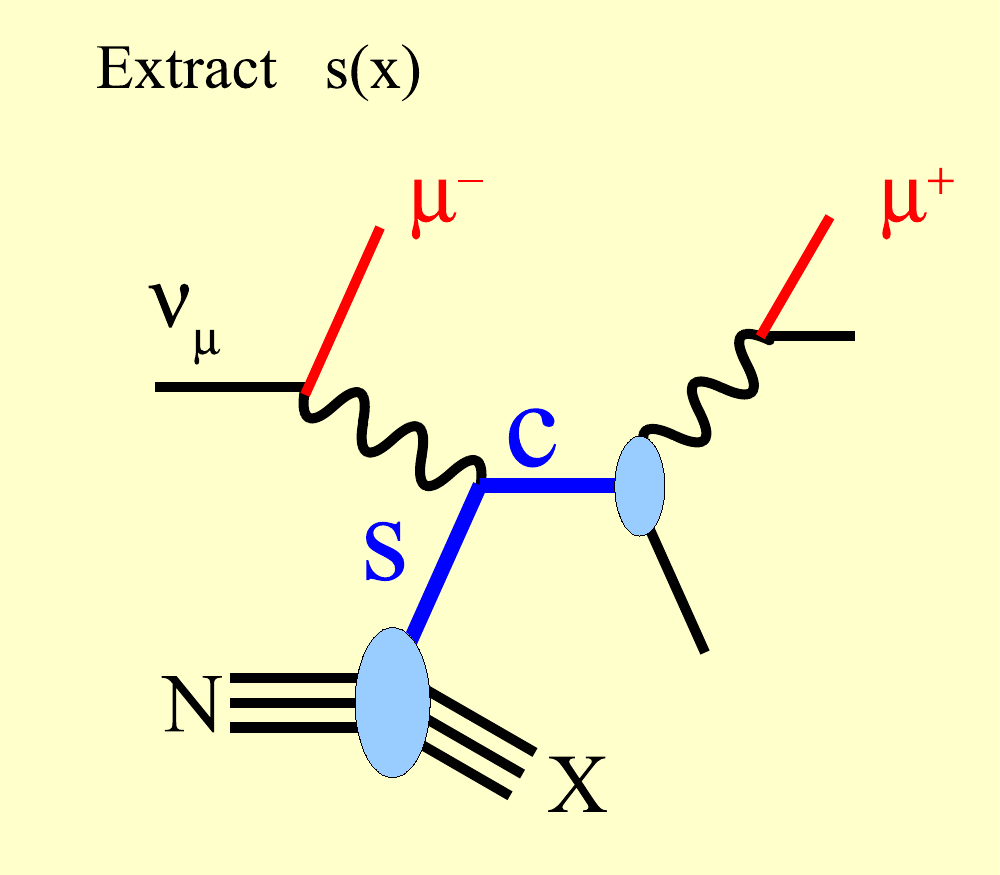}
\hfill{}\includegraphics[clip,width=0.55\textwidth]{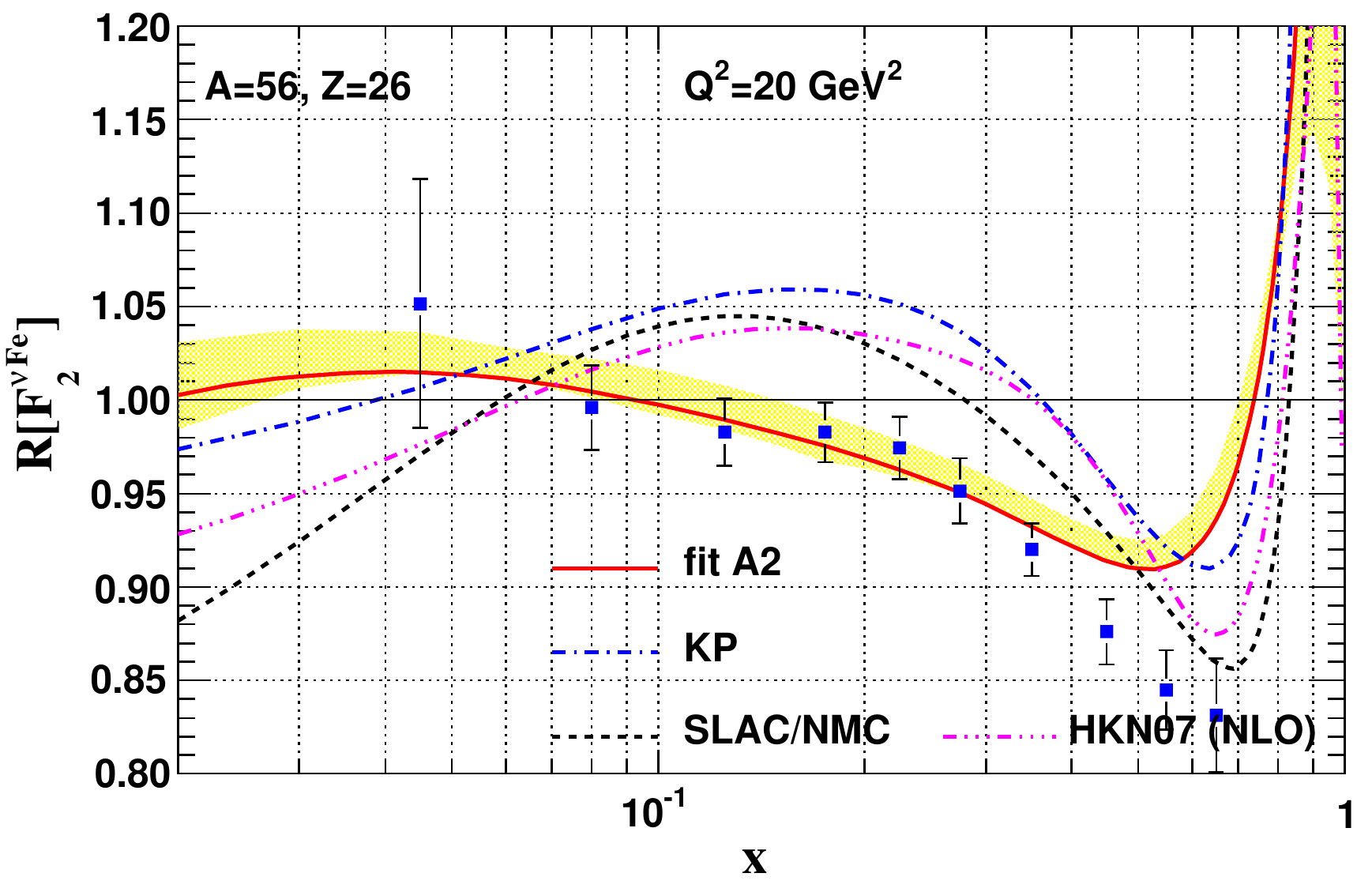}
\par\end{centering}

\caption{{\small{}The computed nuclear correction ratio, $F_{2}^{Fe}/F_{2}^{D}$
as a function of $x$ for $Q^{2}=20\, GeV^{2}$. Figure-a) shows the
basic dimuon process $\nu N\to\mu^{+}\mu^{-}X$. Figure-b) shows the
fit using the $\nu N$ DIS data (fit A2) compared with parameterizations
of the neutral current lepton ($\ell^{\pm}N$) DIS data (KP, SLAC/NMC,
HKN07). The data are from the NuTeV experiment. See Ref.~\cite{Schienbein:2009kk}
for details. \label{fig:nudis}}}
\end{figure}

\begin{figure}
\begin{centering}
\includegraphics[width=0.85\textwidth]{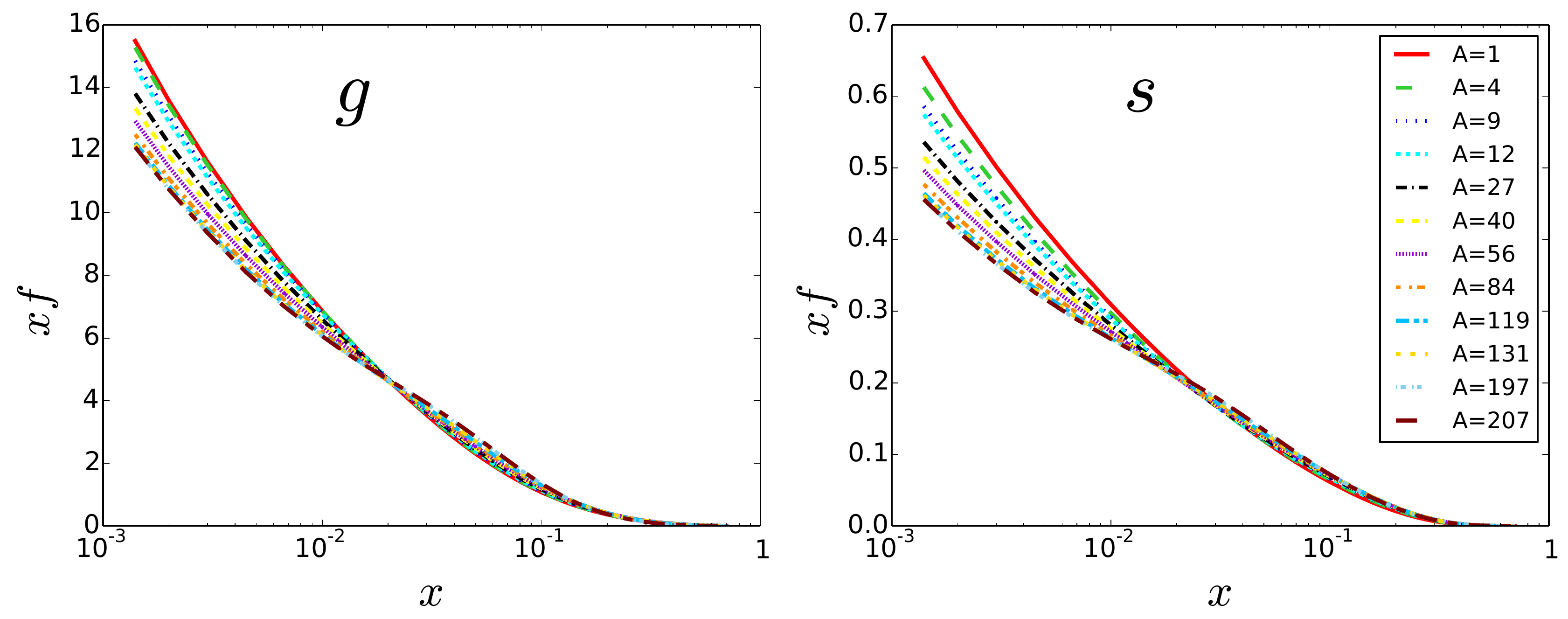}
\par\end{centering}

\caption{{\small{}The nCTEQ15\cite{Kovarik:2015cma} PDFs for selected nuclei
for the gluon and strange PDF at $Q=10\, GeV$. \label{fig:ncteq}}}
\end{figure}

The objective of the nCTEQ project is to obtain the most precise set
of Parton Distribution Functions (PDFs) to facilitate measurements
and interpret hadronic processes at both fixed-target experiments,
HERA, RHIC, Tevatron, and the LHC. The project began when it was realized
that a limiting factor on the proton PDF precision was the nuclear
corrections used for the wealth of nuclei data---particularly the
DIS data which is crucial for flavor differentiation.

As the bulk of the data used in the global analyses of the PDFs comes
from Deeply Inelastic Scattering (DIS) processes, much of this is
measured on heavy targets (e.g., iron or lead) where nuclear corrections
must be taken into account. This data is very important for distinguishing
the separate flavor components in the proton. Surprisingly, the strange
quark PDFs have a large influence on LHC ``benchmark'' processes.

In Fig.~\ref{fig:tevlhc} we note that the heavy quark initiated
contributions ($c\bar{s}$) at the LHC can be 30\% or more of the
total cross section, whereas it is only a few percent at the Tevatron.
Furthermore, the larger $\sqrt{s}$ energy of the LHC probes a much
broader range in rapidity $y$, and hence a broader range in the partonic
$x$. While the LO illustration of Fig.~\ref{fig:tevlhc} is instructive,
in Fig.~\ref{fig:strange} we show the high-precision results of
the NNLO calculation for $\{W^{\pm},Z\}$ using the VRAP program;\cite{Anastasiou:2003ds}
if we are to make full use of this very precise NNLO result, we must
improve the precision of the strange PDF. 

The primary constraint on the strange quark PDF comes from neutrino-induced
DIS dimuon production ($\nu N\to\mu^{+}\mu^{-}X$) on heavy targets.%
\footnote{New data from LHC are beginning to provide information the strange
quark at larger $Q$ and smaller $x$; \textit{cf.} Refs.~\cite{Aad:2012sb,Kusina:2012vh}. %
} Fig.~\ref{fig:nudis}-a) shows the basic dimuon process used to
constrain $s(x)$; the anti-neutrino process can constrain $\bar{s}(x)$.
As the neutrino experiments use heavy nuclear targets (typically iron
or lead), we need to know the nuclear correction to relate this information
back to the proton data. 

Fig.~\ref{fig:nudis}-b) shows the extracted nuclear correction factor
for the neutrino DIS ($\nu N$) processes (fit A2) as compared with
that for charged lepton ($\ell^{\pm}N$) DIS processes (KP, SLAC/NMC,
HKN07), and we observe some significant differences. As was demonstrated
in Ref.~\cite{Kovarik:2010uv}, if we properly incorporate the experimental
correlated errors in the global PDF fit, we are unable to find a nuclear
correction which is compatible with both the $\nu N$ and $\ell^{\pm}N$
data simultaneously;%
\footnote{Note that this difference was present \textit{only} if we imposed
the full constraints of the experimental correlated systematic errors;
if the systematic and statistical errors were added in quadrature,
a common correction factor was obtained. This observation highlights
the importance of the experimental error treatment in the fits, and
resolves a number of questions regarding the compatibility of these
data sets. %
} thus, we must account for this when we extract the strange PDF and
include an additional uncertainty.

\section{The nCTEQ15 PDFs}

\begin{figure}
\begin{centering}
\includegraphics[width=0.85\textwidth]{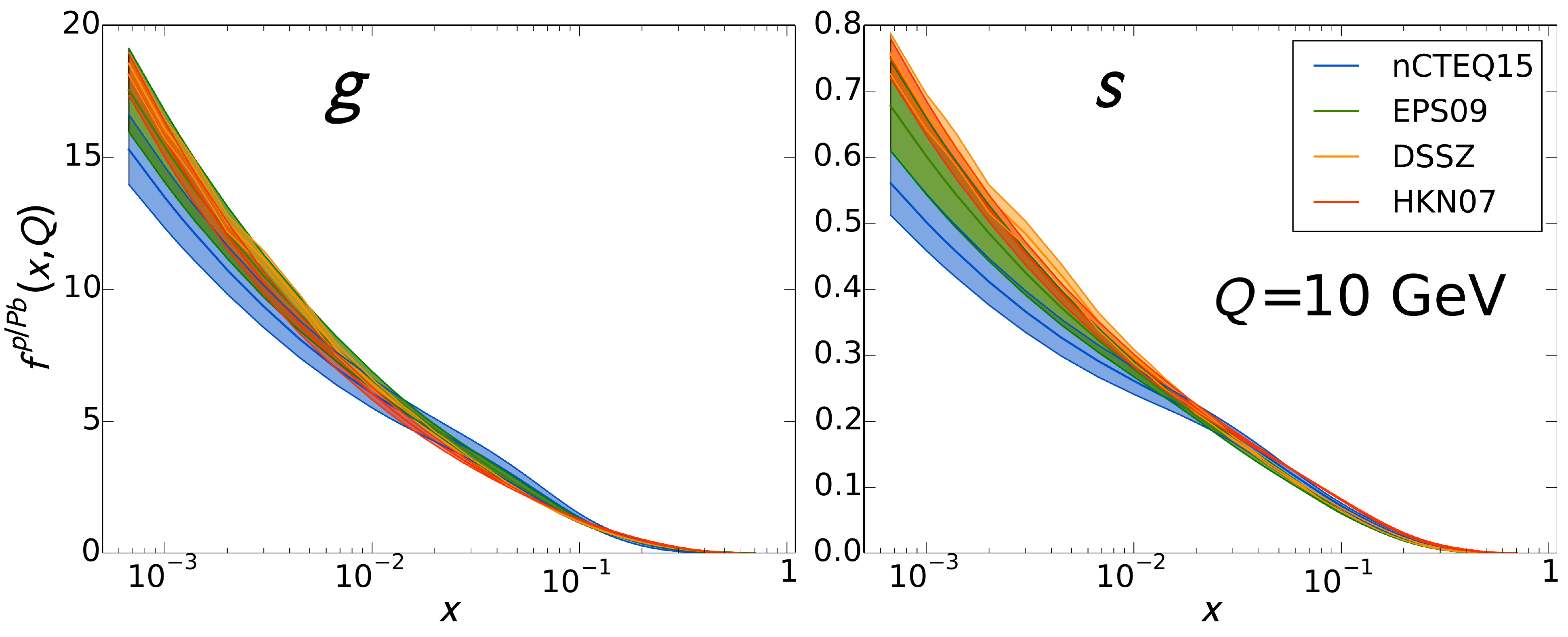}
\par\end{centering}

\caption{{\small{}The nCTEQ15 PDFs showing the uncertainty bands for selected
partons ($g,s)$. For comparison, we also show bands for HKN07,\cite{Hirai:2007sx}
EPS09,\cite{Eskola:2009uj} and DSSZ.\cite{deFlorian:2011fp} \label{fig:bands}}}
\end{figure}

\begin{figure}
\includegraphics[width=0.95\textwidth]{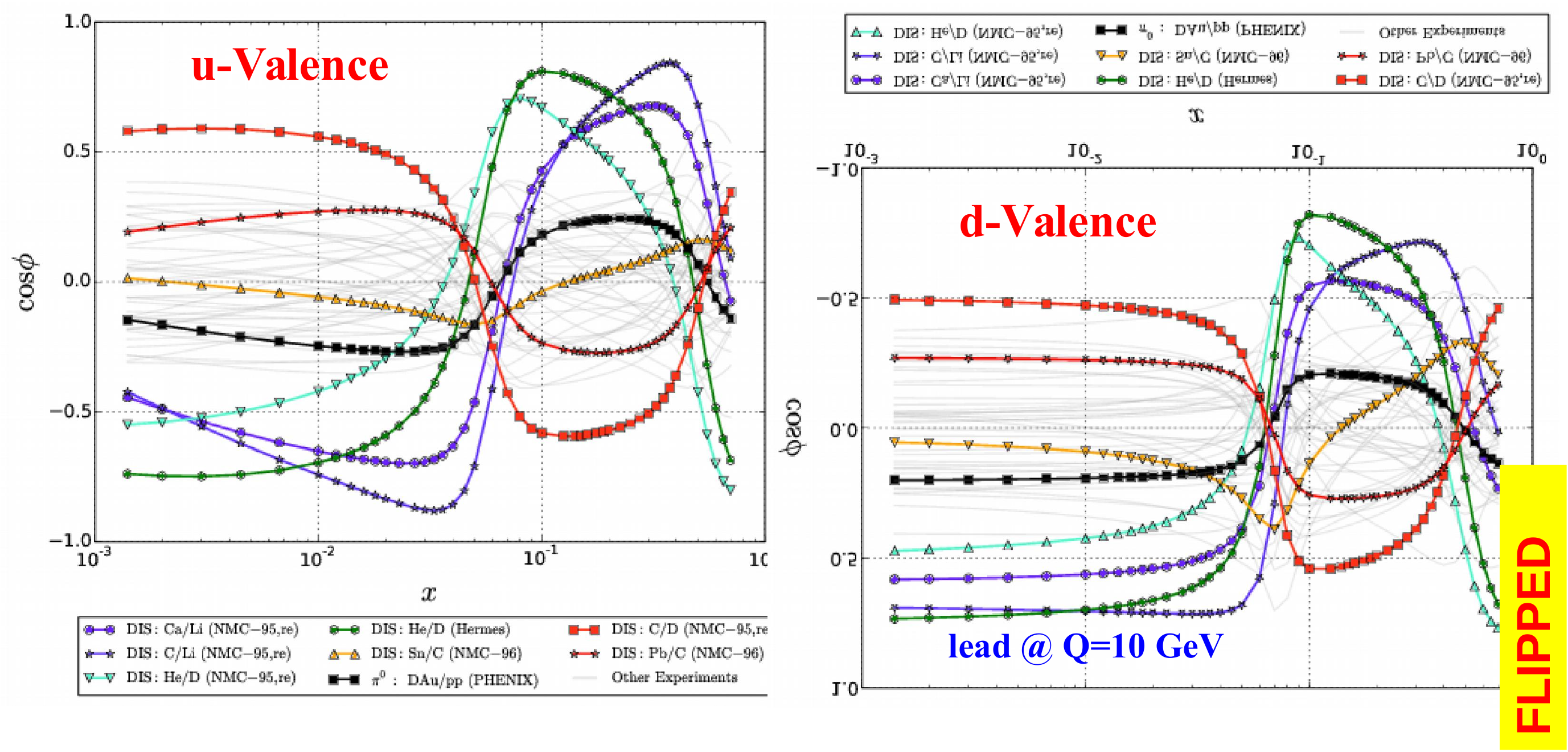}

\caption{{\small{}Correlation measures ($\cos\phi$) for $u_{val}$ (left)
and $d_{val}$ (right) for lead at $Q=10\, GeV$. Eight selected experiments
are highlighted with symbols. To emphasize the anti-correlation between
$u_{val}$ and $d_{val}$ we have flipped the $d_{val}$ plot vertically.
See Ref.~\cite{Kovarik:2015cma} for details. \label{fig:corr}}}
\end{figure}

The nCTEQ framework allows the nuclear correction factors to be integrated
\textit{dynamically} into the fit to better identify tensions between
data sets, and to extract more accurate PDFs when using data from
nuclear targets. We have now released the nCTEQ15 PDFs with error
sets which provide our results of the global analysis for all nuclear
$A$ values.%
\footnote{These are available on-line at the HepForge repository: http://ncteq.hepforge.org/%
}\cite{Kovarik:2015cma} In addition to the Deep Inelastic Scattering
(DIS) and Drell-Yan (DY) processes, we also include inclusive pion
production data to help constrain the gluon PDF. Within our framework
we are able to obtain a good fit to all data. 

Fig.~\ref{fig:ncteq} displays selected partons for a range of nuclear
$A$ values. We have determined the uncertainties using the Hessian
method with an optimal rescaling of the eigenvectors to accurately
represent the uncertainties for the chosen tolerance criteria. In
Fig.~\ref{fig:bands} we compare the nCTEQ15 PDF uncertainty bands
with other sets from the literature. While the general features are
similar, there are some important differences. For example, the nCTEQ15
parameterization allows different correction factors for the up and
down quarks. To investigate which data sets are driving this difference,
we examine the correlation of the data sets with specific flavor components,
and asses the impact of individual experiments. Fig.~\ref{fig:corr}
shows the correlation $\cos\phi$ for the up and down valence as a
function of $x$ for the lead PDFs at $Q=10$~GeV. Selected experiments
are highlighted with symbols. To emphasize the fact that the up and
down valence are relatively anti-correlated, we have vertically flipped
the plot for the down valence to make the correspondence between the
two plots readily apparent. We find that the fit exploits the additional
freedom to reduce the $\chi^{2}$ by an additional $\sim10\%$. While
these are interesting observations, work still remains  to definitively
distinguish parameterization effects from the underlying physics.
In view of the differences, the true nPDF uncertainties should be
obtained by combining the results of all analyses and their uncertainties.

\section{Conclusions}

The nCTEQ15 PDFs represent the first complete analysis of nuclear
PDFs with errors in the CTEQ framework. The framework used for the
nCTEQ15 fit can combine data from both proton and nuclear targets
into a single coherent analysis; thus, it can yield more accurate
PDFs when using data from nuclear targets. 

All in all we find relatively good agreement between different nPDF
sets. Most of the noticeable differences occur in regions without
any constraints from data and so they can be attributed to different
assumptions such as parameterization of the nuclear effects. 

Using the nCTEQ15 fit as a reference, it will be interesting to include
the upcoming LHC data as we continue to investigate the relations
between the proton and the nuclear PDFs.

\pagebreak{}

\Acknowledgments 

I am pleased to thank D.~B.~Clark, 
E.~Godat,
T.~Ježo, 
C.~Keppel, 
K.~Kova\v{r}ík, 
A. Kusina, 
F.~Lyonnet, 
P.~Nadolsky,
J.G.~Morf{\'{i}}n, 
J.F.~Owens, 
I.~Schienbein,
J.Y.~Yu for helpful discussions and collaboration.

\bibliographystyle{utphys}
\bibliography{olness}

\providecommand{\href}[2]{#2}\begingroup\raggedright\begin{thebibliography}{1}

\bibitem{Kovarik:2015cma}
K.~Kovarik {\em et~al.}, ``{nCTEQ15 - Global analysis of nuclear parton
  distributions with uncertainties in the CTEQ framework},''
\href{http://arxiv.org/abs/1509.00792}{{\ttfamily arXiv:1509.00792 [hep-ph]}}.

\bibitem{Kusina:2012vh}
A.~Kusina, T.~Stavreva, S.~Berge, F.~I. Olness, I.~Schienbein, K.~Kovarik,
  T.~Jezo, J.~Y. Yu, and K.~Park, ``{Strange Quark PDFs and Implications for
  Drell-Yan Boson Production at the LHC},''
  \href{http://dx.doi.org/10.1103/PhysRevD.85.094028}{{\em Phys. Rev.}
  {\bfseries D85} (2012) 094028},
\href{http://arxiv.org/abs/1203.1290}{{\ttfamily arXiv:1203.1290 [hep-ph]}}.

\bibitem{Kovarik:2010uv}
K.~Kovarik, I.~Schienbein, F.~I. Olness, J.~Y. Yu, C.~Keppel, J.~G. Morfin,
  J.~F. Owens, and T.~Stavreva, ``{Nuclear corrections in neutrino-nucleus DIS
  and their compatibility with global NPDF analyses},''
  \href{http://dx.doi.org/10.1103/PhysRevLett.106.122301}{{\em Phys. Rev.
  Lett.} {\bfseries 106} (2011) 122301},
\href{http://arxiv.org/abs/1012.0286}{{\ttfamily arXiv:1012.0286 [hep-ph]}}.

\bibitem{Schienbein:2009kk}
I.~Schienbein, J.~Y. Yu, K.~Kovarik, C.~Keppel, J.~G. Morfin, F.~Olness, and
  J.~F. Owens, ``{PDF Nuclear Corrections for Charged and Neutral Current
  Processes},'' \href{http://dx.doi.org/10.1103/PhysRevD.80.094004}{{\em Phys.
  Rev.} {\bfseries D80} (2009) 094004},
\href{http://arxiv.org/abs/0907.2357}{{\ttfamily arXiv:0907.2357 [hep-ph]}}.

\bibitem{Anastasiou:2003ds}
C.~Anastasiou, L.~J. Dixon, K.~Melnikov, and F.~Petriello, ``{High precision
  QCD at hadron colliders: Electroweak gauge boson rapidity distributions at
  NNLO},'' \href{http://dx.doi.org/10.1103/PhysRevD.69.094008}{{\em Phys. Rev.}
  {\bfseries D69} (2004) 094008},
\href{http://arxiv.org/abs/hep-ph/0312266}{{\ttfamily arXiv:hep-ph/0312266
  [hep-ph]}}.

\bibitem{Aad:2012sb}
{\bfseries ATLAS} Collaboration, G.~Aad {\em et~al.}, ``{Determination of the
  strange quark density of the proton from ATLAS measurements of the $W \to
  \ell \nu$ and $Z \to \ell\ell$ cross sections},''
  \href{http://dx.doi.org/10.1103/PhysRevLett.109.012001}{{\em Phys. Rev.
  Lett.} {\bfseries 109} (2012) 012001},
\href{http://arxiv.org/abs/1203.4051}{{\ttfamily arXiv:1203.4051 [hep-ex]}}.

\bibitem{Hirai:2007sx}
M.~Hirai, S.~Kumano, and T.-H. Nagai, ``{Determination of nuclear parton
  distribution functions and their uncertainties in next-to-leading order},''
  \href{http://dx.doi.org/10.1103/PhysRevC.76.065207}{{\em Phys.Rev.}
  {\bfseries C76} (2007) 065207},
\href{http://arxiv.org/abs/0709.3038}{{\ttfamily arXiv:0709.3038 [hep-ph]}}.

\bibitem{Eskola:2009uj}
K.~Eskola, H.~Paukkunen, and C.~Salgado, ``{EPS09: A New Generation of NLO and
  LO Nuclear Parton Distribution Functions},''
  \href{http://dx.doi.org/10.1088/1126-6708/2009/04/065}{{\em JHEP} {\bfseries
  0904} (2009) 065},
\href{http://arxiv.org/abs/0902.4154}{{\ttfamily arXiv:0902.4154 [hep-ph]}}.

\bibitem{deFlorian:2011fp}
D.~de~Florian, R.~Sassot, P.~Zurita, and M.~Stratmann, ``{Global Analysis of
  Nuclear Parton Distributions},''
  \href{http://dx.doi.org/10.1103/PhysRevD.85.074028}{{\em Phys.Rev.}
  {\bfseries D85} (2012) 074028},
\href{http://arxiv.org/abs/1112.6324}{{\ttfamily arXiv:1112.6324 [hep-ph]}}.

\end{thebibliography}\endgroup

\end{document}